\documentclass[aps,prb,twocolumn,superscriptaddress,noshowpacs,a4paper,10pt]{revtex4}
\pdfoutput=1
\usepackage{color}
\usepackage{dcolumn}% Align table columns on decimal point
\usepackage{amssymb}
\usepackage{amsmath}
\usepackage{graphicx}
\usepackage[pdftex]{hyperref}
\hypersetup{colorlinks=true,citecolor=blue,linkcolor=red,urlcolor=blue}
\usepackage[all]{hypcap}

\begin{document}

\author{\firstname{Roberto} \surname{Guerra}}
\email{guerra@sissa.it}
\affiliation{International School for Advanced Studies (SISSA), Via Bonomea 265, 34136 Trieste, Italy}
\affiliation{CNR-IOM Democritos National Simulation Center, Via Bonomea 265, 34136 Trieste, Italy}

\author{\firstname{Erio} \surname{Tosatti}}
\affiliation{International School for Advanced Studies (SISSA), Via Bonomea 265, 34136 Trieste, Italy}
\affiliation{The Abdus Salam International Centre for Theoretical Physics (ICTP), Strada Costiera 11, 34151 Trieste, Italy}

\author{\firstname{Andrea} \surname{Vanossi}}
\email{vanossi@sissa.it}
\affiliation{CNR-IOM Democritos National Simulation Center, Via Bonomea 265, 34136 Trieste, Italy}
\affiliation{International School for Advanced Studies (SISSA), Via Bonomea 265, 34136 Trieste, Italy}

\title{Slider Thickness Promotes Lubricity: from 2D Islands to 3D Clusters}

\begin{abstract}
The sliding of three-dimensional clusters and two-dimensional islands adsorbed on crystal surfaces represent an important test case to understand friction. Even for the same material, monoatomic islands and thick clusters will not as a rule exhibit the same friction, but specific differences have not been explored. Through realistic molecular dynamics simulations of the static friction gold on graphite, an experimentally relevant system,  we uncover as a function of gold thickness a progressive drop of static friction from monolayer islands, that are easily pinned, towards clusters, that slide more readily. The main ingredient contributing to this thickness-induced lubricity appears to be the increased effective rigidity of the atomic contact, acting to reduce the cluster interdigitation with the substrate. A second element which plays a role is lateral contact size, which can accommodate the solitons typical of the incommensurate interface only above a critical contact diameter, which is larger for monolayer islands than for thick clusters. The two effects concur to make clusters more lubric than islands, and large sizes more lubric than smaller ones. These conclusions are expected to be of broader applicability in diverse nanotribological systems, where the role played by static, and dynamic, friction is generally quite important.
\end{abstract}

\maketitle

\section{Introduction}\label{sec.intro}

The distinctive behavior of deposited nano-systems is, mostly due to their large surface-to-volume ratio, often different from that of their constituent parts and from that of the bulk material.\cite{ferrando}
Understanding the diffusion mechanisms and the frictional properties of nano-aggregates of atoms or molecules on surfaces is one of the active research fields in physical science and in tribology .
The forced sliding of artificial metal clusters or of graphene flakes, on "slippery" surfaces such as graphite or graphene is one of the established tools that probe friction at the nanoscale, both experimentally\cite{bardotti,schwartz,schirmeisen,dienwiebel,filippov,carpick,gallani} and theoretically.\cite{landman,lewis,pisov,guerra1, de Wijn, wijk}
Parallel work on atomic scale friction of rare gas adsorbed islands inertially sliding on metal surfaces is also available by Quartz Crystal Microbalance (QCM) experiments,\cite{coffey-krim1996,bruschi,krim1,krim2,mistura_nnano}, and theoretically addressed by molecular dynamics (MD) simulations.\cite{mistura_nnano,cieplak,varini}
These nanoscale sliding phenomena generally involve, for atomically controlled interfaces, the dynamic friction, that is the average force which is necessary to keep the slider in motion.

Static friction -- the force necessary to set a slider in motion from rest -- is another tribological quantity of great importance, although somewhat less frequently addressed.
A large static friction keeps the slider center of mass in a pinned, immobile state; a weak static friction instead permits a facile depinning. The vanishing of static friction, sometimes referred to as "superlubricity", is a peculiar situation realized e.g. by stiff crystalline sliders upon incommensurate crystal surfaces.\cite{revmodphys} For many crystalline islands or clusters adsorbed on good quality crystals, with a contact interface which is stiff and incommensurate, one can actually expect a negligible static friction, thus approaching a tribologically superlubric state.
For an island or a cluster sliding on a such surface,  friction can in fact be argued\cite{mistura_nnano} to consist of a bulk superlubric term, vanishing proportionally to the speed $v$\,$\to$\,$0$, plus an edge term, growing as a sublinear power of the contact area.
The latter does not die out when $v$\,$\to$\,$0$, and contributes significantly to the static friction.\cite{varini} QCM studies\cite{coffey-krim1996, mistura_nnano} confirmed that the dynamic friction of stiff incommensurate rare gas monolayer islands can indeed be very small as expected for such an asymptotically superlubric system.

Metal clusters, many monolayers thick, which have also been pushed by AFM tips to slide on graphite\cite{schwartz}
show  quite similar low friction and sublinear scaling with growing contact area.
\cite{schirmeisen}

\begin{figure}[b!]
  \centering
  \includegraphics[width=\columnwidth]{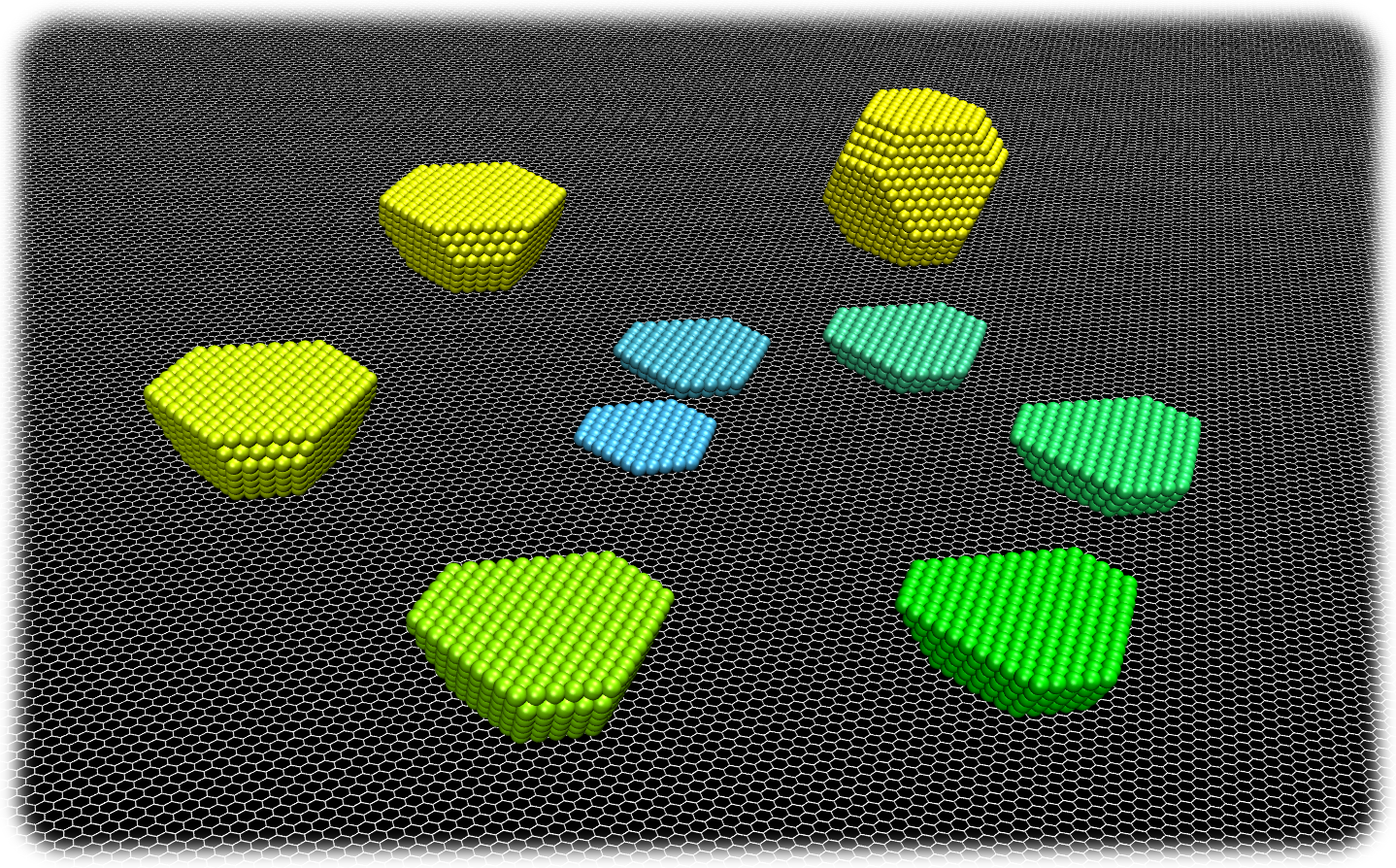}
  \caption{\small A spirally-arranged sequence of gold-on-graphite simulated nanosystems with N$_1$\,=\,90 and increasing number of layers: from 2D island (cyan) to full 3D truncated octahedron Au$_{2075}$ cluster (yellow). }\label{fig.MDsystem}
\end{figure}

To date, no study is available bridging between these two cases, monoatomic islands and thick clusters, a study that would explain what exactly two-dimensional (2D) island and three-dimensional (3D) cluster static friction have in common and what not.
Specifically, if the contact area and everythting else were the same, would lubricity of a cluster be larger, smaller, or the same as that of an island, and why? Moreover, what role would the magnitude of the contact area,  and that of the edge play in the sliding of a cluster as compared to an island of the same material?

Here we provide a first theoretical answer to these questions, by exploring the continuous evolution from one to the other in simulations where an island is made progressively thicker by the addition of crystalline layers. We find that monolayer islands are tendentially more compliant, and therefore more easily pinned, whereas thick clusters are more rigid and more lubric. Moreover, even within the same thickness, static friction depends on the lateral size of the contact. A slippery, lubric contact requires a diameter exceeding a critical value comparable to the incommensurate soliton-soliton spacing.  For Au on graphite that spacing depends on thickness, and it is smaller for a thick cluster than for a monolayer island.

\section{Thickness dependent static friction of islands and clusters }\label{sec.results}

\begin{figure}[b!]
  \centering
  \includegraphics[width=\columnwidth]{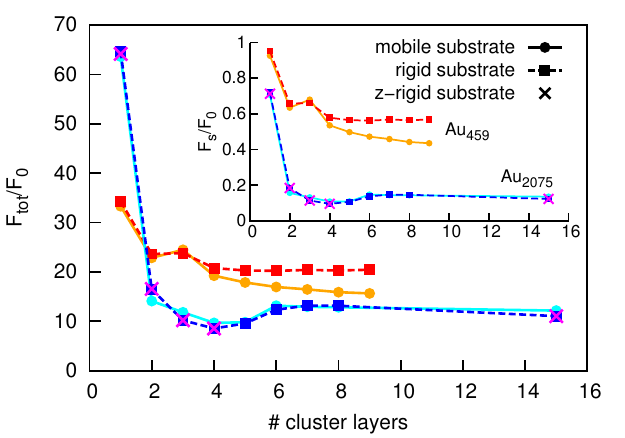}
  \caption{\small Total (main panel) and per-contact-atom (inset) depinning force for N$_1$\,=\,36 (red/orange curve) and N$_1$\,=\,90 (blue/cyan curve) as a function of the number of layers from the 2D island up to the full 3D clusters (N\,=\,459 and 2075, respectively). Solid (dashed) curves correspond to simulations with mobile (rigid) graphite substrate. Crosses indicate x-y only (in-plane) mobile graphite (see text). }\label{fig.Fs_layers}
\end{figure}

We carried out MD simulations of Au nanosliders such as those depicted in Fig.~\ref{fig.MDsystem}, on graphite. The initial slider was a 2D crystalline Au monoatomic island consisting of N$_1$ (ranging from 36 to 90) atoms with a triangular lattice and a hexagonal shape.
The nearest-neighbor Au-Au bonds in the island and C-C bonds in the substrate were both aligned along $y$, and the ($x$,$y$) island center of mass position was that of lowest energy.
Subsequently a second atomic layer with N$_2$ atoms was added on top of the first; then a third layer with N$_3$ atoms on top of the second, and so on, until a full cluster was formed, with N$_1$\,+\,N$_2$\,+\,...\,+\,N$_l$\,=\,N atoms forming a 3D truncated octahedron deposited on an atomistically simulated graphite substrate (see Methods).
Au and C atoms initialized at their respective ideal bulk positions were first allowed to relax at zero temperature. An external $x$-directed driving force $f_{ext}$ was then evenly applied to each Au atom, resulting in a total driving force $F_{ext}=N f_{ext}$, increasing very gradually with time until depinning took place at some threshold value $F_{ext} = F_{tot}$.

We define here the overall static friction force per contact atom as $F_s$\,=\,$F_{tot}$/$N_1$, with $N_1$ the number of Au atoms of the first layer directly in contact with the graphite substrate. For clarity, we will normalize in the following $F_s$ and $F_{tot}$ to the zero temperature depinning $x$-force of a single gold atom, $F^0$\,$\simeq$\,55\,pN, related to the ratio between substrate corrugation and lattice parameter.

Fig.~\ref{fig.Fs_layers} shows (see solid lines) the total depinning force $F_{tot}$/$F_0$ (main panel) and static friction force $F_s$/$F_0$ normalized per contact atom (inset) for two different contact areas, N$_1$ = 36 and 90, as a function of the increasing number of layers $l$, from 2D islands ($l$\,=\,1) up to the development of two full 3D truncated octahedron clusters Au$_{459}$ and Au$_{2075}$ ($l$\,=\,9 and $l$\,=\,15, respectively).

The main result observed was a striking drop of static friction with increasing thickness, an effect stronger for larger planar size. It was found in particular that a second layer ($l$\,=\,2) on top of the contact layer is sufficient to cause a reduction of almost a factor 5, while further added layers have a milder (but not negligible) effect. The modest increase near $l$\,=\,5,6, seen for the larger cluster, arises from from accidentally maximizing the Au-C commensurate regions (colored in green in the subsequent figures) of the solitonic pattern within the size of the interface contact.

At sufficiently low temperatures, the strong ``lubric'' effect of increasing thickness is only modestly influenced by atom mobility in the graphite substrate, as shown by comparison of totally rigid, partly rigid, and fully mobile carbon atoms. It must therefore be attributed to a drop of the contact-induced strain magnitudes at the Au/graphite interface.
In turn, that demonstrates the development of an effective cluster rigidity that increases with thickness $l$, therefore causing a decrease of the slider-substrate lattice interdigitation and a progressive recovery of the natural interface incommensurability expected from the intrinsic Au-C lattice mismatch.

\begin{figure}[b!]
  \centering
{
\setlength{\fboxsep}{0pt}
\setlength{\fboxrule}{0.001pt}
\fbox{\includegraphics[trim=2.3cm 0.6cm 2cm 0.6cm, clip=true, width=0.49\columnwidth]{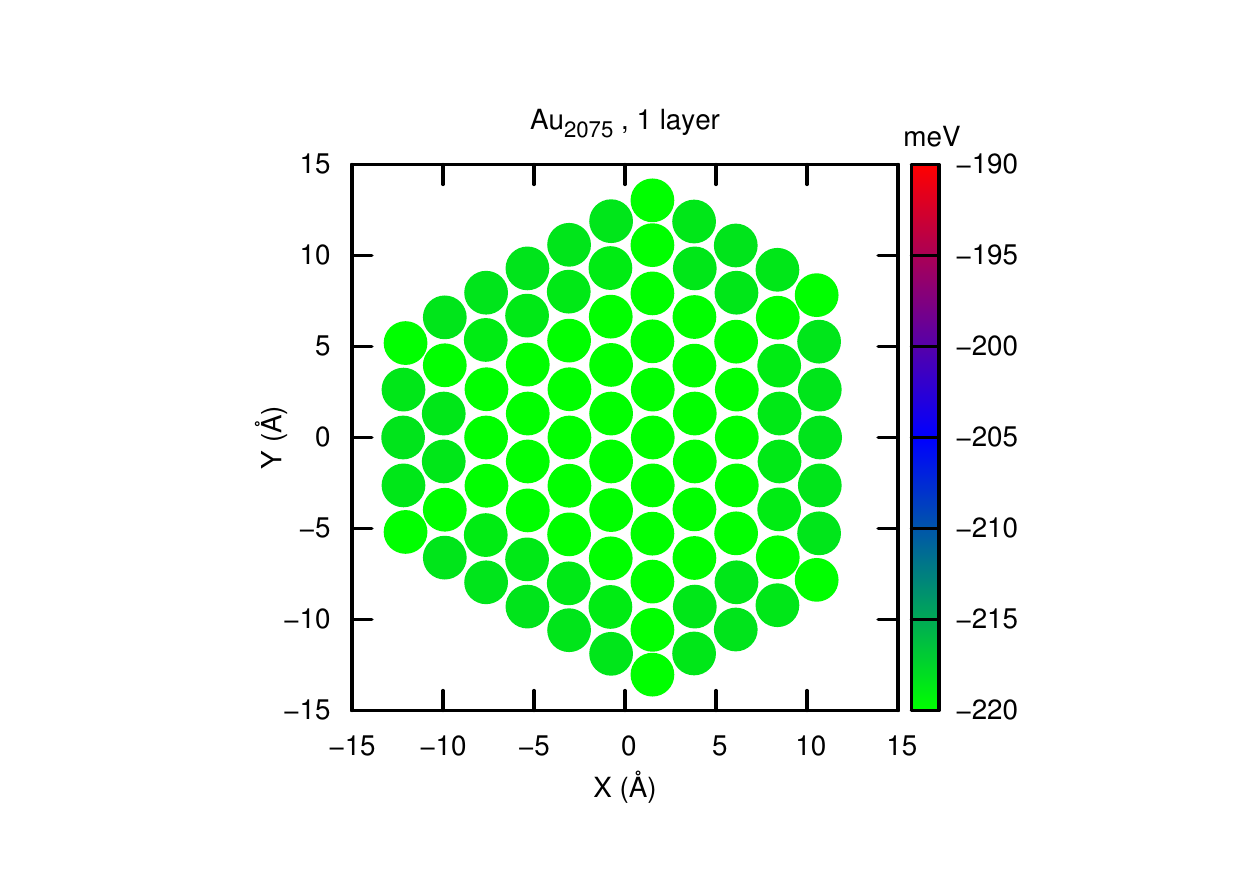}}
\fbox{\includegraphics[trim=2.3cm 0.6cm 2cm 0.6cm, clip=true, width=0.49\columnwidth]{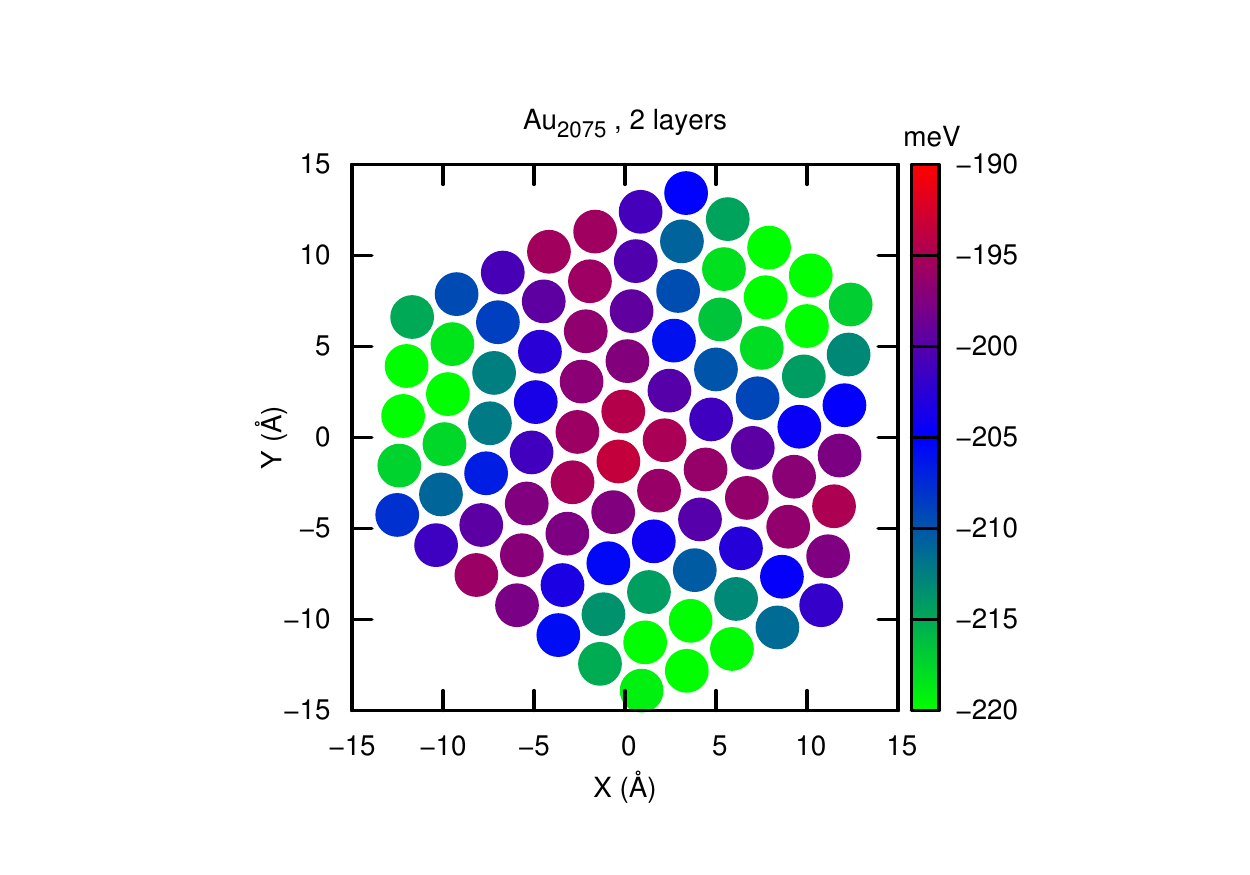}}
}
  \caption{\small Substrate potential map of the 90-atom contact for the fully relaxed, force-free Au-C monolayer island ($l$\,=\,1, left) and bilayer ($l$\,=\,2, right). The monolayer island is soft, and is squeezed into commensurability. The bilayer is stiffer and retains a soliton pattern reminiscent of incommensurability. Due to that, it also develops a spontaneous Novaco-McTague rotation angle of approximately 5.5 degrees.}\label{fig.colormap_islands}
\end{figure}

This inference is directly confirmed in Fig.~\ref{fig.colormap_islands} where a map of the total Au-C potential energy shows a large change between the $l$\,=\,1 2D island (left) and the $l$\,=\,2 3D adsorbate (right). The monolayer Au island lattice behaves as soft, and has deformed into nearly perfect commensurability with graphite. Already in the bilayer, conversely, the soliton pattern, the hallmark of incommensurability between gold and graphite, has formed and is at the origin of the friction drop.
In addition we also note that, unlike the monolayer island lattice which is fully aligned with the substrate lattice, the relaxed bilayer developed a small spontaneous overall rotation angle of about 5.5 degrees. The incommensurability and the higher compressional stiffness of the bilayer are the cause of this characteristic Novaco-McTague rotation,\cite{novaco1,novaco2} whose function is to convert part of the energetically costly misfit compressional stress into a cheaper shear stress. Some very interesting tribological consequences of that rotation were also pointed out in a recent work on 2D colloid friction.\cite{mandelli}
That rotation disappears, at least at the small sizes considered, in the intrinsically softer $l$\,=\,1 monolayer case, where fully aligned adhesion prevails.

These structural findings connect well with the frictional aspects. As is documented in literature for both 1D and 2D systems,\cite{aubry,erdemir} extended interface contacts between stiff incommensurate lattices exhibit a vanishing static friction (superlubricity), whereas those between either commensurate or soft incommensurate lattices always exhibit static friction. The 90-atoms Au monolayer island clearly belongs to the second category. At same size, the bilayer, trilayer, and all our thicker sliders fall in the first category, and actually would be fully superlubric were it not for the boundary atoms, which cause edge pinning as has been discussed in considerable detail for rare gas adsorbed islands.\cite{varini}
For sufficiently extended interface contacts, large enough to develop a solitonic pattern as will be detailed further below, the possibility of this kind of ``Aubry-like'' tribological transition between a pinned soft island and the much more lubric or superlubric multilayer cluster, depicted in our particular system and size, clearly represents a phenomenon of broader interest and relevance than the strict case system studied here.

\begin{figure}[b!]
  \centering
  \includegraphics[width=\columnwidth]{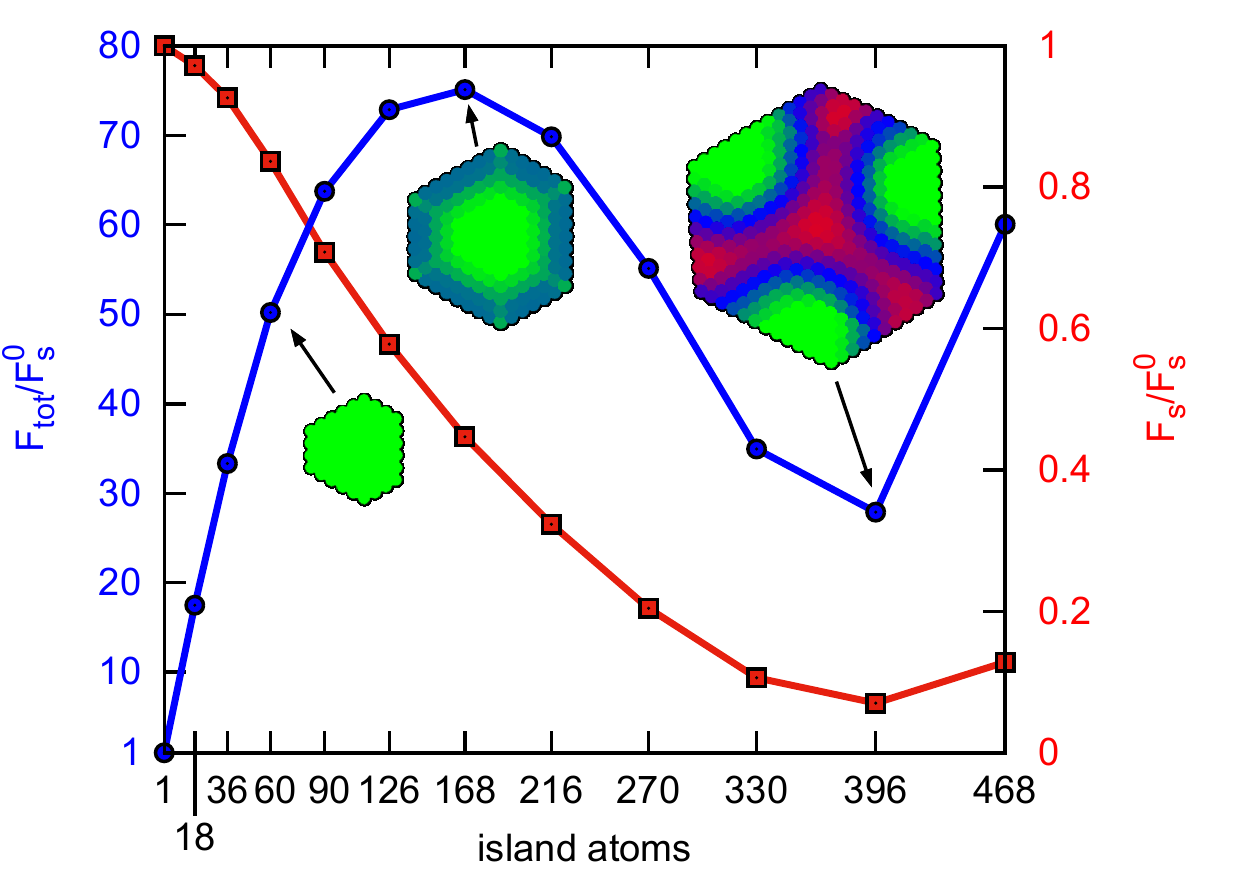}
  \caption{\small Static friction of islands: average depinning force per contact atom ($F_s$) and total depinning force ($F_{tot}$) as a function of the island size. }\label{fig.Fs_1layer}
\end{figure}

\begin{figure}[b!]
  \centering
  \includegraphics[width=\columnwidth]{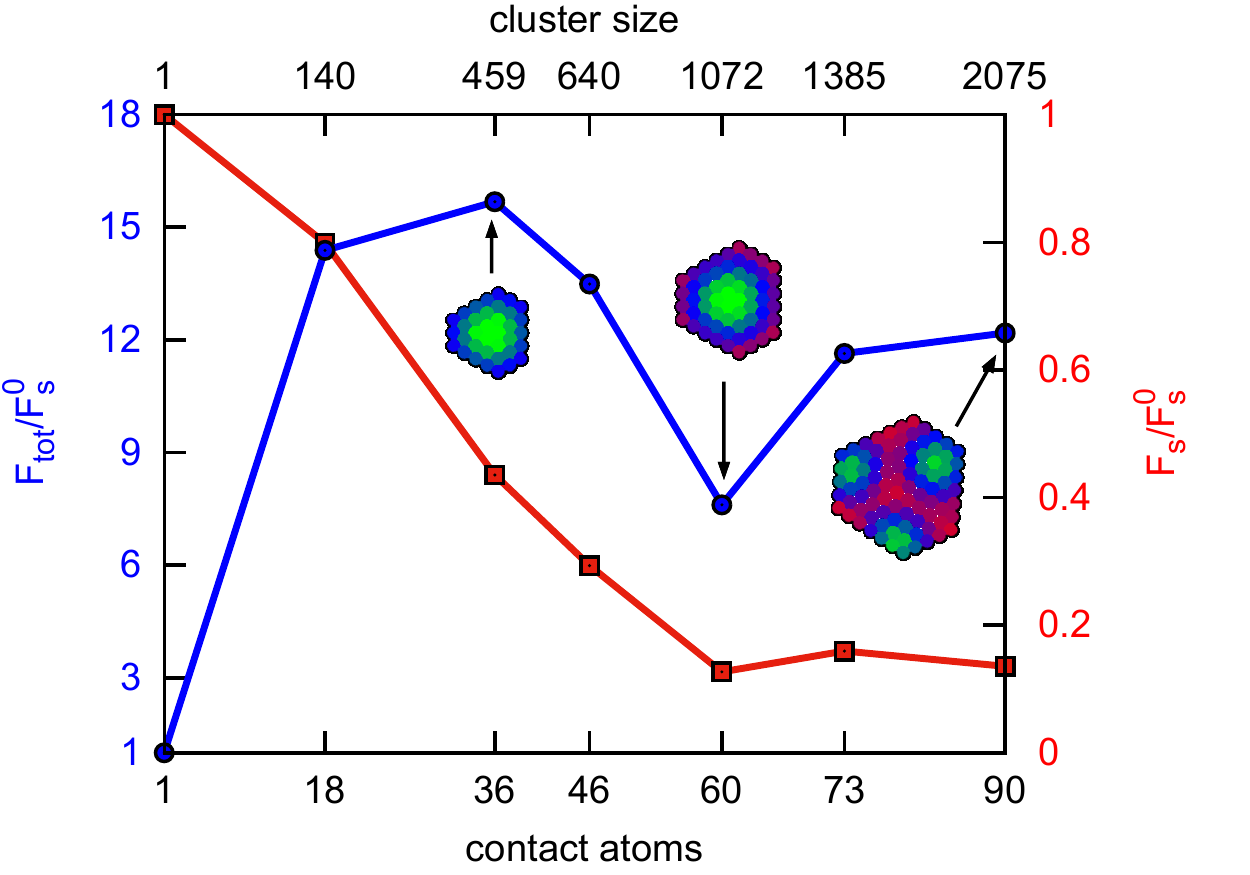}
  \caption{\small Static friction of clusters: average depinning force per contact atom ($F_s$) and total depinning force ($F_{tot}$) as a function of cluster size and contact atoms. }\label{fig.Fs_all}
\end{figure}

\section{Contact area dependent static friction}

Even after normalization per contact atom, the static friction of a nano-slider generally depends on the magnitude of the contact area, chiefly through the size-dependent adsorption geometry.
The adsorption geometry is in turn influenced by two elements: the thickness dependent internal stiffness, just noted; and the natural lattice match or mismatch between the sliding material and the substrate at infinite size. In our case, Au(111) and graphite are mutually incommensurate in bulk. As we just saw the same $N_1$=90 atom contact is stiff, and thus incommensurate and lubric for a thick cluster, but is instead soft, commensurate and pinned as a monolayer island. What will happen for smaller, or for larger, contact areas?
To answer that question, we simulated a sequence of monolayer islands and of thick clusters of increasing size, and extracted their static friction.

Figures~\ref{fig.Fs_1layer}-\ref{fig.Fs_all} show the normalized static friction contact atom $F_s$/$F_0$ and total $F_{tot}$/$F_0$ as a function of the contact atom number $N_1$, separately for 2D monoatomic islands and for 3D clusters. The first result is that at very small size both contacts, Au islands and clusters, are commensurate and aligned with graphite, and therefore strongly pinned. The second result is that both systems turn incommensurate, and much more lubric, at large enough size.
That conclusion is drawn first of all by examining the potential energy maps, with clear soliton features (inserts), as well as by the strongly decreased static friction per atom (red curves). The characteristic contact area where the transition between the two regimes takes place is very thickness dependent, corresponding to about N$_{1c}$\,$\sim$\,400 atoms for islands against N$_{1c}$\,$\sim$\,60 atoms for clusters. The critical contact linear size corresponds to a diameter similar to the soliton-soliton distance, clearly because the incommensurate structure of each case can be realized above but not below that critical area. The soliton  spacing is larger in the monolayer case, and that is the reason why e.g., for N$_1$\,=\,90 the 2D island is still pinned, but the 3D cluster is already lubric.
One further result which we point out is the oscillatory behavior of total static friction, which bounces after the steep drop at N$_1$\,=\,N$_{1c}$. Since the dip was originated by approximate commensurability of the contact diameter and the inter-soliton spacing, that suggests that the subsequent loss and restoration of that commensurability is the source of oscillations. Such oscillations should dampen out at increasing size, upon enhanced sampling of the solitonic pattern.

\begin{figure}[b!]
  \centering
  {
    \setlength{\fboxsep}{0pt}
    \setlength{\fboxrule}{0.001pt}
                          %trim=left bottom right top
    \fbox{\includegraphics[trim=2.3cm 0.6cm 2cm 0.6cm, clip=true, width=0.49\columnwidth]{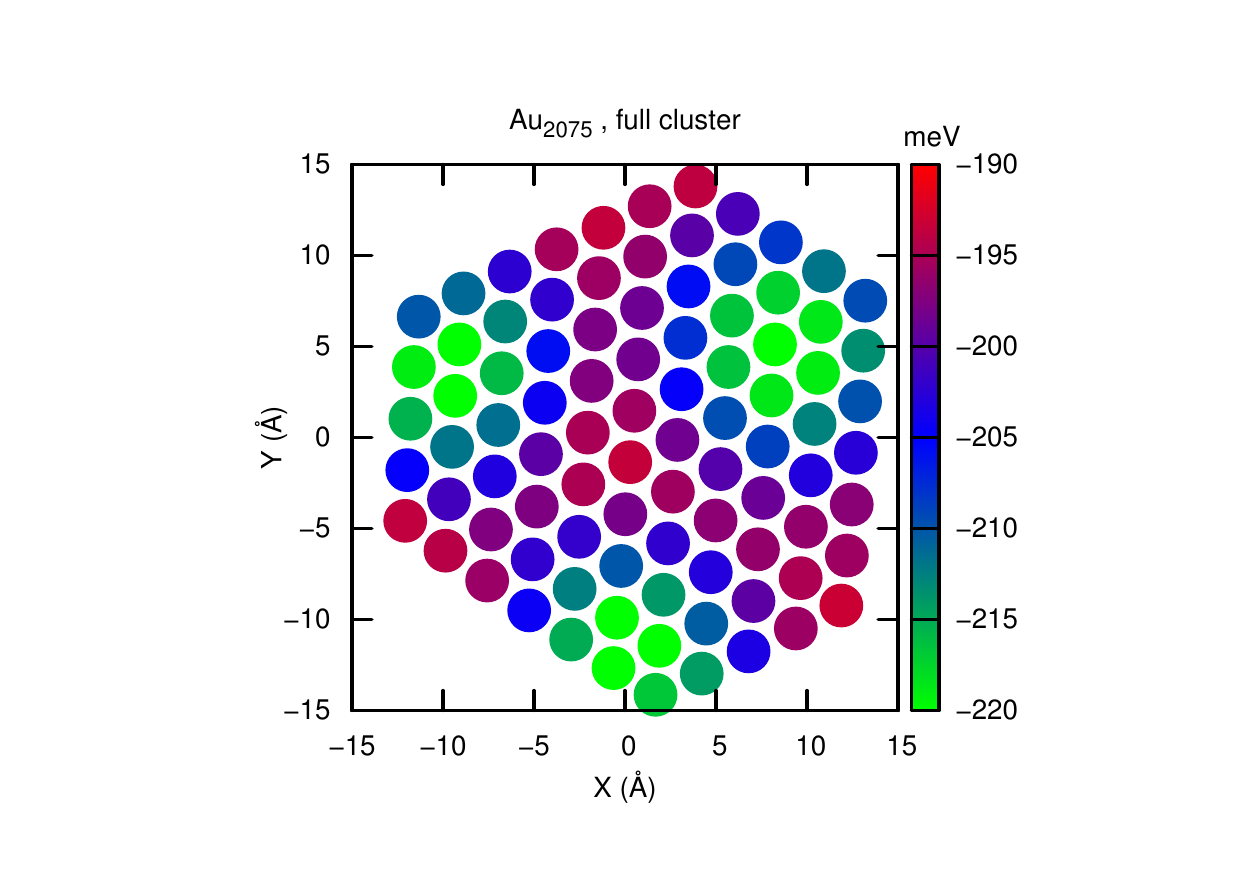}}
    \fbox{\includegraphics[trim=2.3cm 0.6cm 2cm 0.6cm, clip=true, width=0.49\columnwidth]{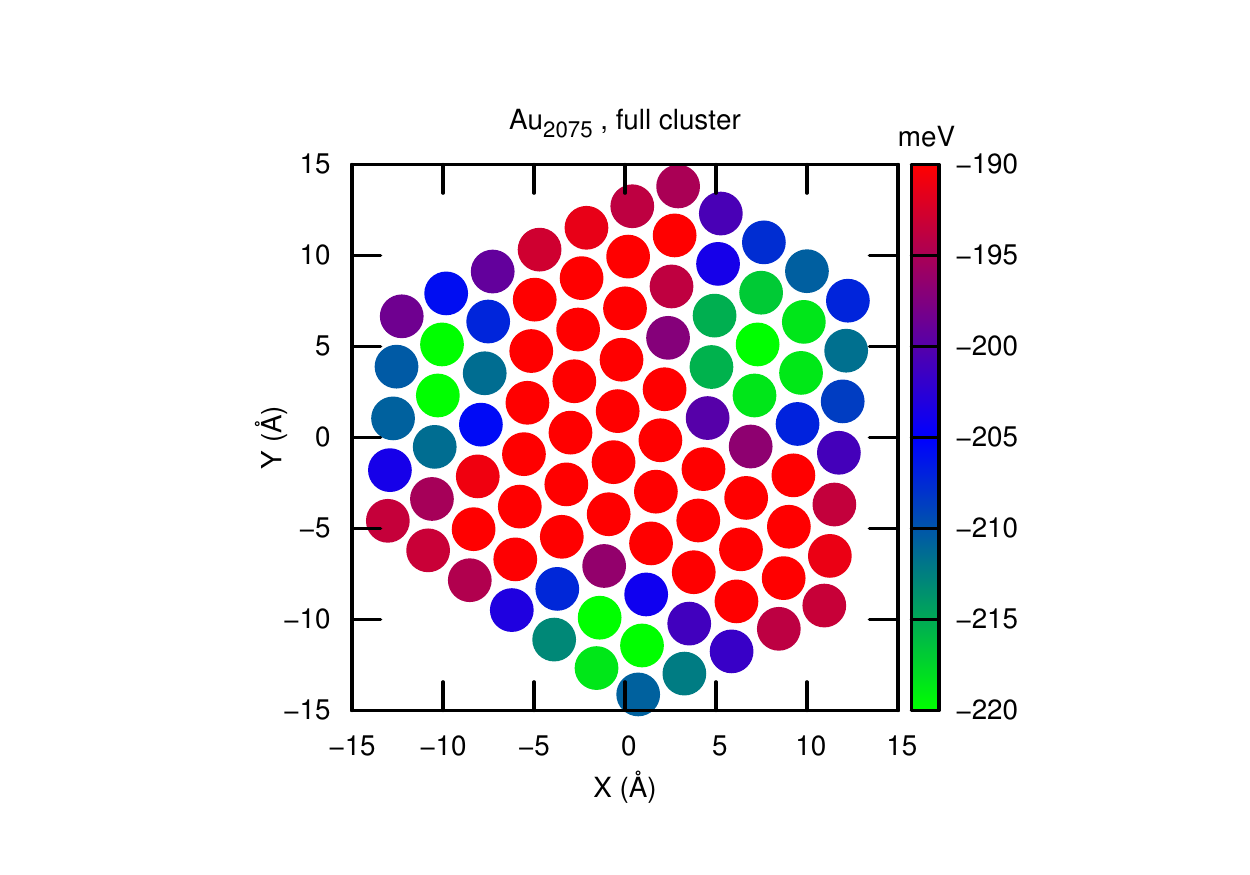}}
  }
  \caption{\small Substrate potential map, highlighting domain-wall patterns, of the contact atoms for the full Au$_{2075}$ relaxed on the mobile (left) and the rigid (right) graphite substrate. }\label{fig.colormap_2panel}
\end{figure}

%\vspace{2mm}

\begin{figure*}[t!]
  \centering
{
\setlength{\fboxsep}{0pt}
\setlength{\fboxrule}{0.000pt}
                      %trim=left bottom right top
\fbox{\includegraphics[trim=2.5cm 0.7cm 2.0cm 1.0cm, clip,width=0.3\textwidth]{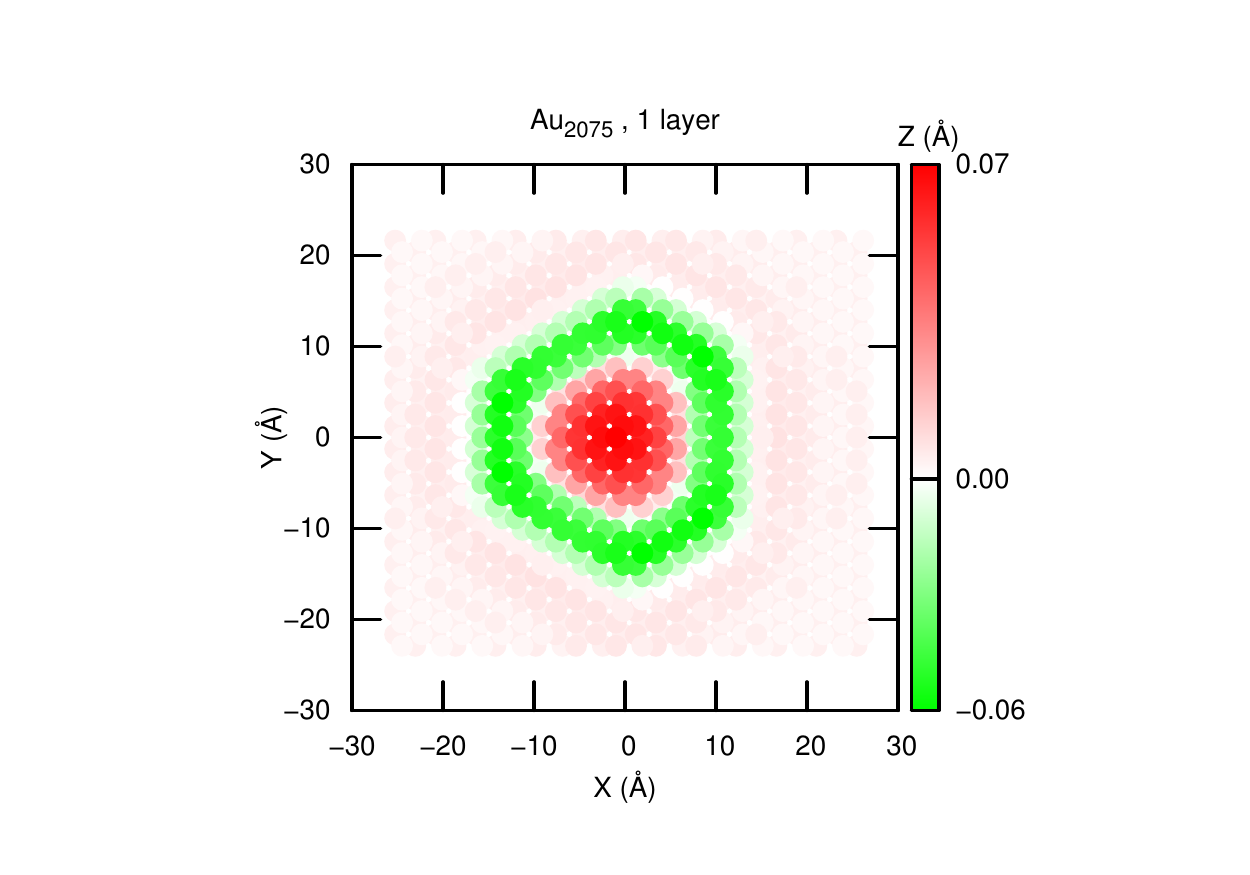}}
\fbox{\includegraphics[trim=2.5cm 0.7cm 2.0cm 0.4cm, clip,width=0.3\textwidth]{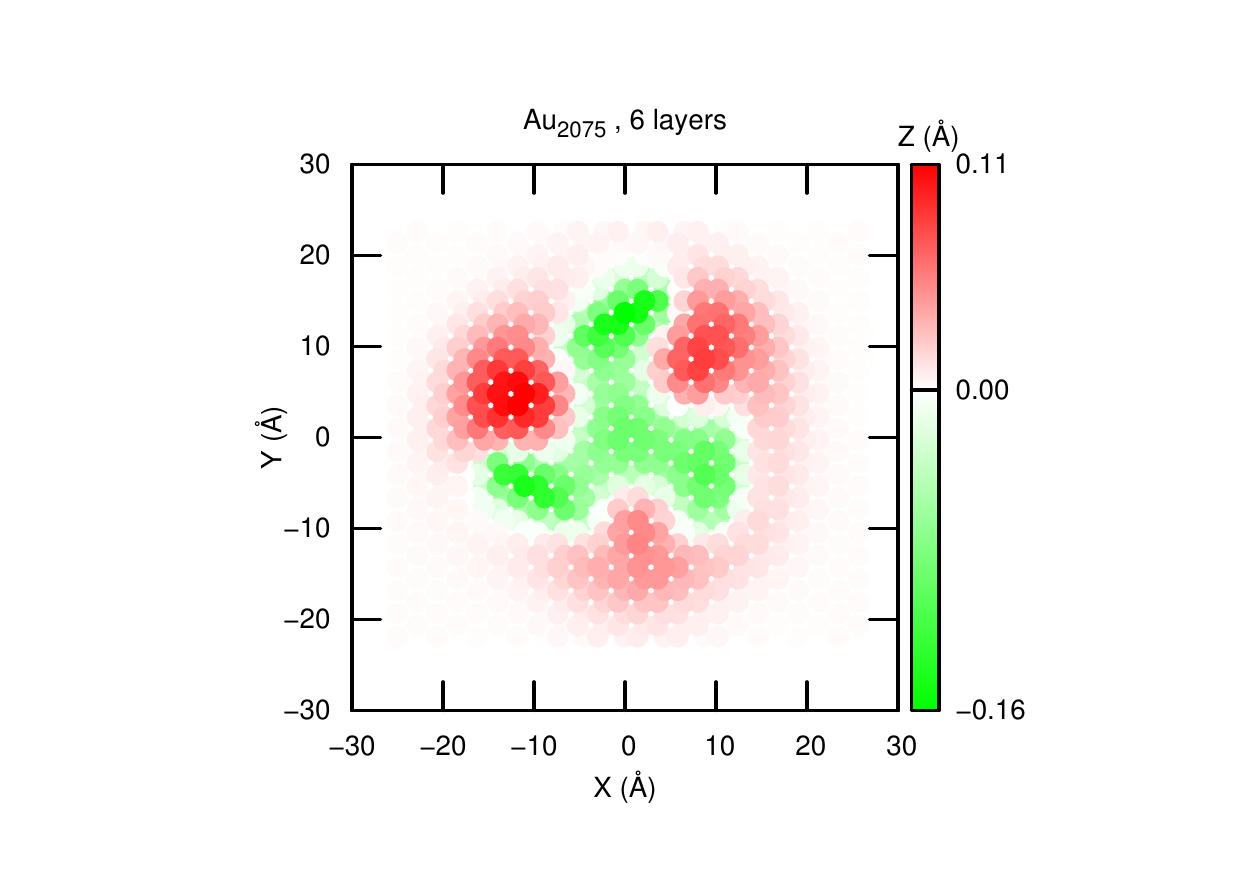}}
\fbox{\includegraphics[trim=2.5cm 0.7cm 2.0cm 0.4cm, clip,width=0.3\textwidth]{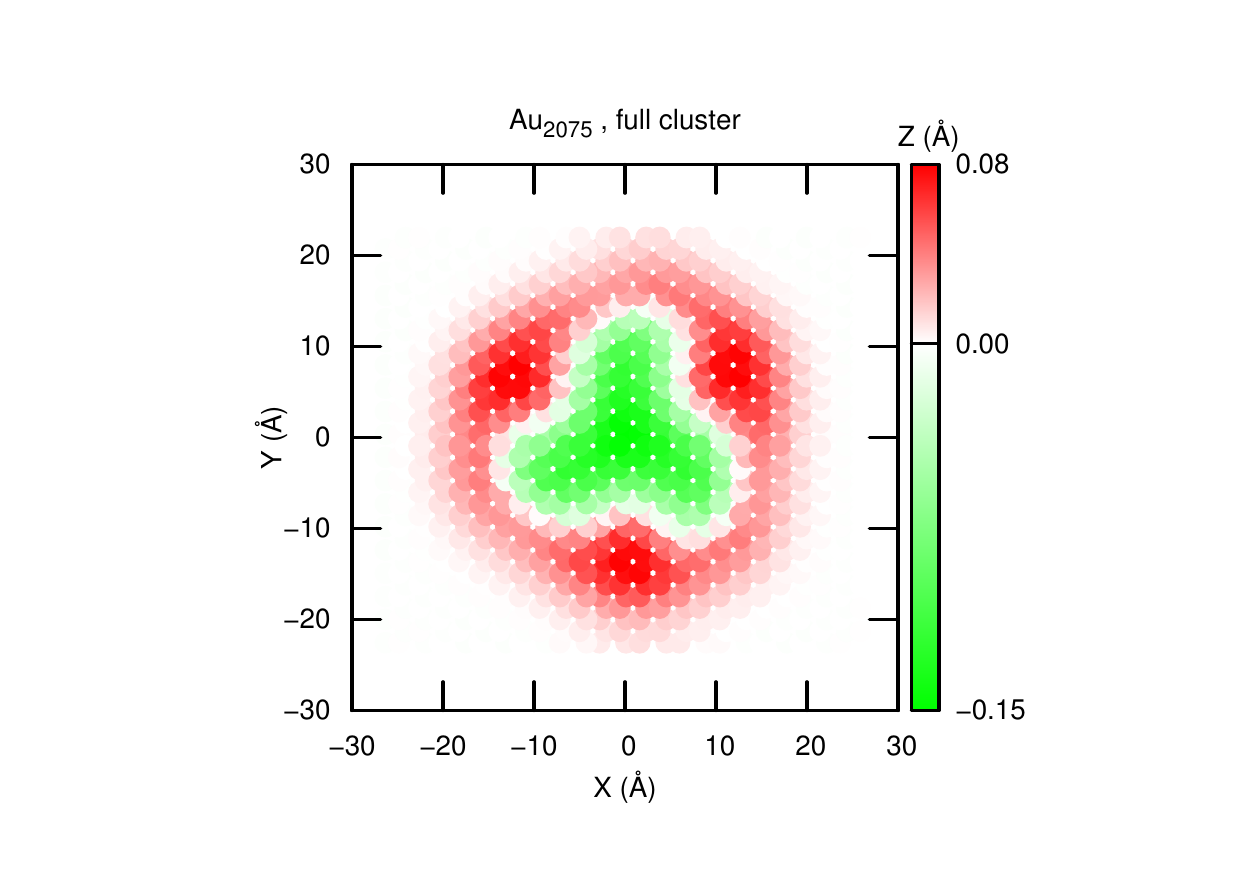}}
}
  \caption{\small $z$-value of top graphite atoms in case of~$l$\,=\,1 (left), $l$\,=\,6 (center), and maximum $l$ (right) for the deposited Au$_{459}$ Au$_{2075}$ cluster relaxed on a mobile graphite with no external force applied. The color scale reports the minimum, the average, and the maximum $z$-value. }\label{fig.zmap_graphite}
\end{figure*}

\section{Graphite substrate deformability}

Besides the influence of the contact layer stiffness, the observed trend of the depinning force with thickness and contact area can depend, to some extent, on the substrate deformability, i.e. the capability of the latter to rearrange in order to minimize the stress forces. While intra-plane carbon rearrangements are strongly limited by the huge Young modulus of graphite, out-of-plane displacements involve very soft modes and, therefore, are likely to play some role.

That question is addressed in Figure~\ref{fig.Fs_layers} where, in addition to the fully mobile graphite substrate (circles), we also show the adsorbate $F_s$ and $F_{tot}$ values obtained by depinning simulations over an artificially rigid graphite in the bulk configuration (squares), and also over a graphite surface whose carbon atoms were allowed to move only along the $x$-$y$ plane (crosses), the $z$-motion artificially frozen. The near coincidence of the latter two results indicates that the in-plane displacements of graphite is, as expected, negligible.
Interestingly, we note that the rigid substrate tends, compared to the fully mobile one, to increase the static friction of the smaller clusters while decreasing that of the larger ones. The latter behavior is usually expected in the case of incommensurate interface geometries, since a rigid substrate does not contribute in the screening of the interfacial stress, thus favoring depinning. On the other hand, one expects an opposite behavior in nearly commensurate interfaces, where any structural rearrangements allowed by a mobile surface would inevitably reduce the pinning potential.\cite{guerra0}  These expectations are well brought out by our Au$_{459}$ case, where a reduced (increased) force is required to depin the cluster deposited over a mobile (rigid) substrate.

The effect of the substrate mobility on the Au contact atoms in terms of substrate potential energy can be observed by the direct comparison of the two panels of Figure~\ref{fig.colormap_2panel}. Clearly, for the large contact area of the Au$_{2075}$ cluster, the rigidity of the substrate enhances the solitonic pattern, with a consequent increased system lubricity.

For the sake of completeness we have reported in Fig.~\ref{fig.zmap_graphite} the $z$-displacement of the mobile graphite substrate, for the deposited island ($l$\,=\,1, left), the partial cluster ($l$\,=\,6, center), and the full cluster (right) of the Au$_{2075}$ system.
We note that the out-of-plane displacement of mobile graphite (to be compared with the average Au-C distance, $d$\,$\simeq$\,2.73\,\AA), maps the soliton pattern at the Au interface so that greater (smaller) interfacial separations are produced for interfacial Au atoms located on top (hollow) sites.
In all cases, we observe a wrinkling effect in the graphite surface below (and around) the cluster contact area, which could in principle affect cluster motion.
A particular geometry of these surface wrinkles may, e.g., locally alter the distance between the mating surfaces, thus influencing the effective contact area.
The precise value of the measured friction force will be the result of a complex competition between all the discussed processes.
A similar interface phenomenon, in the form of driving-induced substrate puckering, has been recently suggested to explain the dissipation mechanism in graphene-based dynamical friction experiments.\cite{carpick,gallani}

\section{Discussion and Conclusions}\label{sec.conclusions}

On a crystal surface, the friction -- more precisely the static friction -- of a monolayer island and that of a three-dimensional cluster are not the same, even for the same material and the same contact area. The cluster turns, in general, to be elastically stronger, the island softer. As a result, islands have a stronger tendency to become pinned, whereas clusters are more lubric. We have presented realistic simulations illustrating these conclusions for the test case of Au on graphite, here used as a demonstration case.
On a theoretical basis, based on Aubry's theory\cite{aubry}, the superlubricity of infinite crystals in contact requires two distinct, yet complementary, ingredients: interface incommensurability and stiffness.
For nanoscale sliders, a smooth transition between pinned (large friction) and lubric (low friction) adsorbate systems can be realized, at least conceptually, in two manners. The first manner is by evolving from a softer 2D monolayer to bilayer, trilayer, and so on, until reaching a full 3D structure, hopefully developing the sufficient interface stiffness. In the case presented here, the transition took place already between the monolayer and the bilayer. The second manner is by increasing the lateral size of a stiff-enough nanocontact. This will turn from pinned to lubric as soon as its size gets large enough to accommodate the natural (due to lattice or orientational mismatch) solitonic pattern, thus sampling the required interface incommensurability. In general the critical size will be thickness dependent, small for a thick cluster, and largeer for a monolayer island.

Experimentally, monoatomic rare gas islands are realized and inertially pushed in Quartz Crystal Microbalance (QCM) experiments at submonolayer coverage.\cite{coffey-krim1996,bruschi,krim1,krim2,mistura_nnano} In this technique the islands cannot be seen, and due to strong wetting their thickness cannot be manipulated to form clusters.

Gold and antimony clusters sliding on graphite have been extensively studied and pushed laterally by AFM tips, determining their dynamic friction. In principle this technique should be able to measure their static friction as well. Assimilating static and (low velocity) dynamic friction, one can surmise that a clear lubric behavior and ready sliding has often been observed for crystalline Au clusters.\cite{schwartz,schirmeisen}
The size-dependent friction in these cases was satisfactorily rationalized by models assuming rigid crystalline clusters. Our present results identify thickness-induced stiffening of the cluster as the underlying reason why the assumption of rigidity can stand. In the hypothetical case of a Au monolayer island the rigidity assumption would be a lot less warranted, and lubricity should be reduced, particularly for small contact sizes.

More generally, the role of thickness and of contact size revealed here may be relevant to novel ways of controlling friction in nano-materials.

\section{Methods}\label{sec.method}

Both the 2D islands and the 3D truncated octahedron face-centred-cubic clusters were generated starting from a bulk-gold slice, and their shape optimized with an empirical embedded-atom-method (EAM) potential.\cite{johnson} The Au-C interaction was described by a Lennard-Jones potential, parametrized by $\sigma$\,=\,2.74\,\AA, $\varepsilon$\,=\,22\,meV, and a cutoff radius of 7\,\AA,\cite{lewis} which are known to yield realistic corrugation barriers of the graphite substrate against cluster dynamics.
All simulations generally started at the lowest energy equilibrium configuration (position and angle) of the cluster on the substrate. We verified that, when starting from different initial positions, the clusters tended to find the same positioning on graphite right after the application of a small external force.
Our goal being the relative evolution of static friction with thickness, we restricted to the applied force direction along x.  The absolute static friction will of course somewhat  depend
upon the force direction, there is no reason to expect that these relative variations would.
The dynamics of a two-layer and a three-layer fully mobile graphite substrate is considered and described by an optimized Tersoff intra-layer potential\cite{tersoff} and inter-layer Kolmogorov Crespi registry dependent potential RDP0.\cite{kolmogorov_crespi}
Although more elaborate potentials are available for a description of the planar dynamics of graphite, they are not necessary here, because we are mostly concerned with statics, and because the in-plane graphite rigidity is anyway very large.
We checked that two or three mobile graphite layers gave sufficiently similar results to mimic a semi-infinite bulk graphite.
Periodic boundary conditions are applied along the $x$-$y$ directions of each graphite plane, which is formed by up to 36$\times$64\,=\,2304 carbon atoms.
Static friction measurement is obtained at $T$\,=\,0 by increasing the external force -- evenly applied to all the atoms -- at quasi-adiabatic steps of 1\,meV/\AA, with a threshold of 10$^{-6}$\,\AA/ps in the center-of-mass velocity.
We point out at the outset that the zero temperature is chosen for a precise conceptual reason: this is the only state in which static friction is well defined and nonzero for a nanoscale size slider. Unlike infinite sliders that do not diffuse thermally, a nanoscale island or cluster will always diffuse given a sufficiently long time scale  making static friction strictly
zero at all finite temperatures (this is sometimes referred to as ``thermolubricity'').

\vspace*{0.5 cm}

\subsection*{Acknowledgments}

\noindent This work was partly funded by the Swiss National Science Foundation through SINERGIA contract CRSII2\_136287, by the ERC Advanced Grant No.\ 320796-MODPHYSFRICT, and by COST Action MP1303.

% \subsection*{Author Contributions}
% 
% \noindent R.G., E.T., and A.V. developed the theoretical model, performed the simulations, and contributed to the numerical data analysis. All authors discussed the results and contributed to the writing of the manuscript.
% 
% 
% \subsection*{Competing Financial Interests}
% 
% \noindent The authors declare no competing financial interests.

\end{document}